\documentclass[fleqn,usenatbib]{mnras}

\usepackage{newtxtext,newtxmath}

\usepackage[T1]{fontenc}
\usepackage{ae,aecompl}
\usepackage[utf8]{inputenc}

\usepackage{graphicx}	
\usepackage{amsmath}	
\usepackage{float}

\title{}
\title[L$_V$ vs. P$_{orb}$ for Stellar-Mass Black Holes]{Relating Peak Optical Luminosity and Orbital Period of Stellar-Mass Black Holes in X-ray Binaries}

\author[Blackmon \& Maccarone ]{
Victoria A. Blackmon $^{1,2}$
 and Thomas J. Maccarone $^{1}$\\
$^{1}$Department of Physics \& Astronomy, Texas Tech University, Lubbock TX, 79410-1051, USA\\
$^{2}$Department of Physics \& Astronomy, West Virginia University, Morgantown WV, 26056, USA\\
}

\date{Accepted XXX. Received YYY; in original form ZZZ}

\pubyear{2022}

\begin{document}
\label{firstpage}
\pagerange{\pageref{firstpage}--\pageref{lastpage}}
\maketitle

\begin{abstract}
We compare the peak optical luminosity with the orbital period for a sample of 22 stellar-mass  black  hole  candidates  with  good  measurements  of  both  quantities.  We find that the peak absolute magnitude for the outbursts follows a linear  relation with {$M_{V,peak}=3.48 (\pm 0.85)-3.89 (\pm 0.91) {\rm log}P_{orb}$}, which corresponds to a \textbf{$L_V \propto P_{orb}^{1.56\pm 0.36}$} power law relation. Excluding V4641~Sgr which is a strong outlier and not likely to have outbursts produced by the standard disc instability model, in addition to BW Cir and V821~Ara\textemdash \; both of which have highly uncertain distances; the new correlation for the 19 sources is found to be $M_{V,peak}= 3.01\;(\pm 0.93)-3.21\;(\pm 1.04){\rm log}P_{orb}$, which corresponds to $L_V \propto P_{orb}^{1.28 \pm 0.42}$. This is an analogous relationship to the ``Warner relation" between orbital period and peak luminosity found for cataclysmic variables. We discuss the implications of these results for finding black hole X-ray binaries in other galaxies and in our own Galaxy with the Large Synoptic Survey Telescope and other future large time domain surveys.

\end{abstract}

\begin{keywords}
Black hole physics -- X-ray: binaries 
\end{keywords}



\section{Introduction}

Relatively common in our Galaxy, stellar mass black holes (BHs), varying anywhere from 3 $M_{\odot}$ to a bit over 20 $M_{\odot}$, are the result of core collapses of massive stars. Only about 60 of them make up the currently known population of stellar mass black holes in X-ray binary systems (XRBs) in the Milky Way (\citealt{Corral_Santana_2016}). Nearly 24 years ago, a project (\citealt{Shahbaz_1998}) was carried out using 8 soft X-ray transients\textemdash\; a sub-class of low mass X-ray binary systems (LMXBs) known to undergo varying levels of low energy X-ray emission with extended periods of quiescence \textemdash\; to find a correlation between their optical outburst amplitudes (i.e. the change in magnitude) and their orbital periods. Here, we discuss the implications of a similar relationship regarding the peak optical luminosity and orbital period of a sample of 22 stellar mass black holes in X-ray binary systems with known distance, outburst magnitude, and orbital period.

We observe the way in which the behavior of these black hole binaries follows the typical trend that objects with longer orbital periods possess intrinsically brighter outbursts  than those with shorter periods (\citealt{Cannizzo_1998}; \citealt{Wu_2010}; \citealt{Portegies_Zwart_2004}). We also note the relevancy of this relationship and its involvement in future black hole binary system research as the extent of wide field time domain optical sky surveys increases with goals to provide wider fields of observation and improved precision in addition to the ability to explore previously unobserved phenomena (\citealt{Djorgovski_2013}). As we approach the completion of the Large Synoptic Survey Telescope (LSST) we aim to possess the advanced capability to discover new transients first in the optical band.   Already, in the past few years, a few new black hole X-ray binaries have been discovered first from their optical outbursts in shallower surveys (\citealt{Yao_2021}; \citealt{Tucker}).  We expect that with the discovery of {\it extragalactic} black hole X-ray binaries, we can revisit our proportionality model using larger sample sets with two overarching objectives: 

\begin{enumerate}
\item to further strengthen our drawn relationship\textemdash which will consequently provide us with a stronger understanding of what to expect in regard to, both, detection in the optical and data analysis

\item to bring attention to potential outliers observed to appear as stellar-mass black holes in X-ray binaries which, upon further analysis, may not obey expectations

\end{enumerate}


\section{Observational Data}

We have plotted the relationship between the absolute magnitude and period for the objects in our sample set utilizing the magnitude-distance relation and corresponding data from the stellar-mass black hole catalog, BlackCAT (\citealt{Corral_Santana_2016}), with the data collected in early 2019, in Figure 1 [with the difference being whether V4641~Sgr, V821~Ara (which is also known as GX~339-4) and BW Cir are included]. We determined the absolute magnitudes of each black hole XRB using the existing distance estimate. We converted to $V$ magnitudes for systems with some other band for the peak measurement by using the filter-specific apparent magnitude constants for $B$, $R$, and $J$ bands from the Vega-AB conversion table in \cite{Blanton}. These values were then subtracted from each absolute magnitude to yield the corrected values.
Furthermore, we accounted for reddening and extinction in these calculations by adding a correction term $A_V= 3.1E(B-V)$ (\citealt{Cardelli}),
where color excess values for each stellar mass black hole were taken from BlackCAT's tabulations.%

One of the objects, BW~Cir, shows a very different distance estimate in the data from {\it Gaia DR2} than the earlier data in BlackCAT (\citealt{Ghandi}). The {\it Gaia} parallax requires an unphysically small donor star, and hence is probably due to chance superposition of a foreground star.  Notably, the source's parallax and proper motion change substantially between DR2 and EDR3. We also note that V821~Ara, while very well studied, has only lower limits for its distance.

After plotting the log of the orbital period against our calculated absolute magnitudes for the original 22-candidate sample, we derived a linear fit following the equation:
\begin{equation} \label{eq1}
M_{V,peak}=3.48 (\pm 0.85)-3.89 (\pm 0.91) {\rm log}P_{orb} \;\;\;\;  
\end{equation}
which corresponded to the power law relation
\begin{equation}
L_V \propto P^{1.56\pm0.36}
\end{equation}
shown in the top panel of Figure 1, with standard error accounted for. Furthermore, we obtain a correlation coefficient of $-0.79$ between absolute magnitude and orbital period for the given sample, with the sign being negative because of the convention that magnitudes are more negative for brighter objects.

Some of the scatter in the data is likely due to inclination angle variations, but this is likely limited to about one magnitude of the $\approx 6$ magnitude range seen.  The observed dynamically confirmed black hole X-ray binaries are in a range of inclinations of about 40-70 degrees.  The lower limit comes from the difficulty of making mass estimates for systems without measurable ellipsoidal modulations, and the upper limit probably comes from the fact that the flared outer accretion disk intercepts the X-rays for edge-on systems, making them hard to detect in outburst with all-sky X-ray monitors.  Within that range, a factor of about 2 in brightness difference is expected due to inclination angle effects, consistent with what has been seen in a study of novalike cataclysmic variables with well-understood inclinations \citep{Howell}.


\section{Discussion}

In observing BH outbursts in the X-ray band, we expect to see higher X-ray luminosities as periods increase from object to object represented as 
\begin{equation}
    L_X \propto P_{orb} M
\end{equation}
which follows from the relationships in \cite{Portegies_Zwart_2004} and \cite{Wu_2010} that show that the peak luminosity in Eddington units is proportional to the orbital period. This relation saturates at the Eddington luminosity for sources with periods longer than about 20 hours, but relatively few of the sources in our sample exceed this period substantially. Further, following \cite{vanParadijs}:  

\begin{equation}
L_V \propto L_X^{1/2}R
\end{equation}
where $R$ represents the outer radius of a given black hole's accretion disk, we are further able to describe proportionality between the peak optical luminosity and orbital period. 

Then, taking Kepler's third law in a convenient form, we have:
\begin{equation}
\label{eq6}
a= 1 {\rm AU}\left(\frac{M_{tot}}{M_\odot}\right)^{1/3}\left(\frac{P_{orb}}{\rm yr}\right)^{2/3},
\end{equation}
where $a$ is the binary semi-major axis, $M_{tot}$ is the total binary system mass, and $P_{orb}$ is the orbital period.
  
 As these BHs accrete gas from their companion stars, under the assumption that this infalling gas has a net angular momentum, we find that the circulating gas contains enough angular momentum to produce a circular orbit with radius denoted by $r_{circ.}$, the circularization radius (\citealt{Rosas-Guevara}). For the black hole X-ray binaries, the ratio of the donor mass to accretor mass is always small, and the ratio of the circularization radius $r_c$ and the orbital axis $a$, will be in a small range near 0.2, so it is a good approximation to assume $R\sim{a}$ (\citealt{Frank}).
 
 Under the assumption that these values are proportional, relation (5) can be regarded as
 \begin{equation}
     L_V \propto L_V^{1/2}a
 \end{equation}
showing proportionality between the optical luminosity of each of our black hole candidates and the product of their X-ray luminosities and semi-major axes of orbit.
Going one step further by substituting equation (7) into relation (8), it is shown that
\begin{equation}
    L_v \propto P_{orb}^{1/2}\cdot M_{total}^{1/3}\cdot P_{orb}^{2/3}
\end{equation}
where we insert $P_{orb}$ for $L_X$, based on our primary relation (see: $M_V$ vs. $log \Sigma$, \citealt{vanParadijs}).

Given the narrow range in black hole mass distribution for low mass X-ray binaries (\citealt{Ozel}; \citealt{Farr_2011}), a result of their specific evolutionary paths, along with the fact that the donors are all much lighter than the accretors so that the total mass of the system is approximately equal to the black hole mass, our proportionality reduces as follows:
\begin{equation}
      L_{V,peak} \propto P^{7/6}.
\end{equation}

Our final result corresponds to the relation between peak optical luminosity and orbital period, which we find to be the X-ray binary analog to that observed for cataclysmic variables (CVs) described by the Warner relation (\citealt{Warner}), who finds the relation between $M_V$ and $P$ for dwarf novae with distance determinations as follows:
\begin{equation}
    M_V=5.64\;(\pm 0.13)-0.259\; P(hr)\;(\pm 0.024) \;\;\;\;\;\;\;\;\;
\end{equation}

However, with one of our primary goals being to distinguish potential outliers in our analysis, we observed a particularly bright magnitude for object V4641 Sgr with an orbital period of 1.83 and an absolute magnitude of $-6.25$. With an absolute magnitude nearly twice as bright as objects of similar periods, we acknowledge two characteristics that may help to describe the reason this object does not appear to correlate closely with the $L_x \propto P_{orb}$ relation:

\begin{enumerate}
    \item  V4641~Sgr has a massive, luminous, blue, late B-type secondary star (\citealt{MacDonald_2014}), allowing for this binary to be considered a high mass X-ray binary system (HMXB) by many despite undergoing its mass transfer process via Roche-lobe overflow, commonly associated with LMXBs.
    
    \item As a result of its secondary star, V4641~Sgr may have a different mechanism responsible for its outbursts (\citealt{MacDonald_2014}) relative to that of other low mass black hole X-ray transients (BHXTs).
    
    \item Some indications also exist that the jet in this source is beamed toward the Earth (\citealt{Orosz}), so it may be the case that the optical emission is primarily from a beamed jet.
    
\end{enumerate}

Since we have reason to believe V4641 Sgr is quite different from the rest of our sample, we determined the fit equation, excluding this candidate in addition to the aforementioned V821 Ara and BW Cir, to be:
\begin{equation}
M_{V,peak}= 3.01\;(\pm 0.93)-3.21\;(\pm 1.04){\rm log}P_{orb}
\end{equation}
as shown in the bottom panel of Figure 1 solely for the purposes of comparison. We obtain a slightly worse correlation coefficient, -0.72, after excluding these candidates, indicating that some fraction of the correlation is driven by the fact that V4641~Sgr is abnormally bright, and also at long period, but the slope is still significantly different from zero at the 3$\sigma$ level, and this version is likely to be more representative of the global population of soft X-ray transients.  It is equivalent to a relation such that $L_{V,peak}\propto{P_{orb}^{1.28 \pm 0.42}}$, which is much closer to the analytic approximation $L_{V,peak}\propto{P_{orb}}^{7/6}$ than the relation including V4641 Sgr, V821 Ara and BW Cir.
Regarding the goodness of fit for linear fits previously described, $M_{V,peak}$ uncertainties were determined from the use of
\begin{equation}
    2 \cdot\;(\frac{\sigma_{D}}{D})
\end{equation}
where $\sigma_{D}$ and $D$, respectively, are the distance error and source distances given in Table 1. In conjunction with other parameters from Table 1, the reduced chi square for the 19 stellar-mass BH candidates was 25.56 for 17 degrees of freedom. 

In the coming years, we expect that with the completion of the Large Synoptic Survey Telescope, in particular, locating more than the currently known stellar-mass black holes may be possible in our own galaxy and will greatly increase the possibility of finding extragalactic black hole XRBs of this same size-class, as the LSST's projected goal is to survey the entire available sky from the Cerro Pachón mountain in Chile (\citealt{Ivezic}) every few nights. With the ability to detect BH outbursts first in the optical band, the likelihood of observing this in the nearest galaxy to the Milky Way, the Andromeda Galaxy (M31), with a survey telescope like the LSST would require us to consider factors like our current ability to detect outbursts in optical, the distance of M31 from Earth (assuming ground-based detection), object visibility in the quiescent state and \textemdash arguably, most importantly\textemdash\;instrument sensitivity limits.  

Generally speaking, regular observations of these black hole XRBs must be conducted (often by robotic telescope monitoring, e.g. \citealt{Russell_2019}) in order to accumulate accurate optical detections of the initial onset of X-ray outbursts.

Based on the estimates made using the Faulkes monitoring system described in \cite{Russell_2019}, 15 out of a sample of 17 LMXB outbursts were likely to be detected in the optical band prior to X-ray detection 80-90\% of the time. If we were to consider a sample of 22 BHs in M31 at a distance of $770$kpc, we might expect that, by the same estimate, about 50\% of these objects would peak at a magnitude brighter than $V=24$.  LSST is expected to reach 24th magnitude for individual exposures (although that limit may be somewhat harder to reach in the optically brightest parts of nearby galaxies).  With sufficiently high cadence, then, these transients may be detectable at the distance of M31, and the longer period systems such as GX~339-4, V404~Cyg and XTE~J1859+226 would be detectable out to several megaparsecs with LSST-like sensitivity.  Other time domain surveys which make detecting such transients a priority, like BlackGEM and ZTF, could be competitive with LSST by using longer exposures and/or stacking exposures taken over a few days worth of time.  Crucially, these could then allow LSST or other optical surveys to be used to trigger X-ray observations (with e.g. Swift for the brightest and most nearby galaxies, or more sensitive telescopes for more distant galaxies), and radio observations, when the ngVLA starts to operate.


\section{Conclusions}

We have presented the finding of a relation between peak optical luminosity and orbital period for a sample of known stellar-mass black holes in X-ray binaries which fits reasonably well to analytic predictions. This analysis will become applicable to new data as more binaries like these may be discovered in nearby galaxies. By fitting the orbital periods and absolute magnitudes of a sample of stellar mass black holes linearly, it is confirmed quantitatively that longer periods yield more luminous absolute magnitudes, though we recall the presence of V4641 Sgr which we consider an outlier in our data due to its abnormally bright magnitude given its value for orbital period.

\section{Acknowledgments}
We would like to thank Phil Charles, Poshak Gandhi, Erik Kuulkers, Dave Russell, Tariq Shahbaz and Jorge Casares Vel\'azquez for valuable discussions.

\section{Data Availability}
The data underlying this article are compiled by \cite{Corral_Santana_2016} in BlackCAT at \url{https://www.astro.puc.cl/BlackCAT/}.
\bibliographystyle{mnras}
\bibliography{bibliography} 

\begin{thebibliography}{}
\makeatletter
\relax
\def\mn@urlcharsother{\let\do\@makeother \do\$\do\&\do\#\do\^\do\_\do\%\do\~}
\def\mn@doi{\begingroup\mn@urlcharsother \@ifnextchar [ {\mn@doi@}
  {\mn@doi@[]}}
\def\mn@doi@[#1]#2{\def\@tempa{#1}\ifx\@tempa\@empty \href
  {http://dx.doi.org/#2} {doi:#2}\else \href {http://dx.doi.org/#2} {#1}\fi
  \endgroup}
\def\mn@eprint#1#2{\mn@eprint@#1:#2::\@nil}
\def\mn@eprint@arXiv#1{\href {http://arxiv.org/abs/#1} {{\tt arXiv:#1}}}
\def\mn@eprint@dblp#1{\href {http://dblp.uni-trier.de/rec/bibtex/#1.xml}
  {dblp:#1}}
\def\mn@eprint@#1:#2:#3:#4\@nil{\def\@tempa {#1}\def\@tempb {#2}\def\@tempc
  {#3}\ifx \@tempc \@empty \let \@tempc \@tempb \let \@tempb \@tempa \fi \ifx
  \@tempb \@empty \def\@tempb {arXiv}\fi \@ifundefined
  {mn@eprint@\@tempb}{\@tempb:\@tempc}{\expandafter \expandafter \csname
  mn@eprint@\@tempb\endcsname \expandafter{\@tempc}}}

\bibitem[\protect\citeauthoryear{{Blanton} \& {Roweis}}{{Blanton} \&
  {Roweis}}{2007}]{Blanton}
{Blanton} M.~R.,  {Roweis} S.,  2007, \mn@doi [\aj] {10.1086/510127}, \href
  {https://ui.adsabs.harvard.edu/abs/2007AJ....133..734B} {133, 734}

\bibitem[\protect\citeauthoryear{Cannizzo}{Cannizzo}{1998}]{Cannizzo_1998}
Cannizzo J.~K.,  1998, \mn@doi [The Astrophysical Journal] {10.1086/305123},
  493, 426

\bibitem[\protect\citeauthoryear{{Cardelli}, {Clayton}  \& {Mathis}}{{Cardelli}
  et~al.}{1989}]{Cardelli}
{Cardelli} J.~A.,  {Clayton} G.~C.,   {Mathis} J.~S.,  1989, \mn@doi [\apj]
  {10.1086/167900}, \href
  {https://ui.adsabs.harvard.edu/abs/1989ApJ...345..245C} {345, 245}

\bibitem[\protect\citeauthoryear{Corral-Santana, Casares, Mu{\~{n}}oz-Darias,
  Bauer, Mart{\'{\i}}nez-Pais  \& Russell}{Corral-Santana
  et~al.}{2016}]{Corral_Santana_2016}
Corral-Santana J.~M.,  Casares J.,  Mu{\~{n}}oz-Darias T.,  Bauer F.~E.,
  Mart{\'{\i}}nez-Pais I.~G.,   Russell D.~M.,  2016, \mn@doi [Astronomy {\&}
  Astrophysics] {10.1051/0004-6361/201527130}, 587, A61

\bibitem[\protect\citeauthoryear{Djorgovski, Mahabal, Drake, Graham  \&
  Donalek}{Djorgovski et~al.}{2013}]{Djorgovski_2013}
Djorgovski S.~G.,  Mahabal A.,  Drake A.,  Graham M.,   Donalek C.,  2013, in ,
  Planets, Stars and Stellar Systems.
Springer Netherlands, pp 223--281, \mn@doi{10.1007/978-94-007-5618-2_5}, \url
  {https://doi.org/10.1007%2F978-94-007-5618-2_5}

\bibitem[\protect\citeauthoryear{Farr, Sravan, Cantrell, Kreidberg, Bailyn,
  Mandel  \& Kalogera}{Farr et~al.}{2011}]{Farr_2011}
Farr W.~M.,  Sravan N.,  Cantrell A.,  Kreidberg L.,  Bailyn C.~D.,  Mandel I.,
    Kalogera V.,  2011, \mn@doi [The Astrophysical Journal]
  {10.1088/0004-637x/741/2/103}, 741, 103

\bibitem[\protect\citeauthoryear{Frank, King  \& Raine}{Frank
  et~al.}{2002}]{Frank}
Frank J.,  King A.,   Raine D.,  2002, Accretion Power in Astrophysics, 3 edn.
Cambridge University Press, \mn@doi{10.1017/CBO9781139164245}

\bibitem[\protect\citeauthoryear{Gandhi, Rao, Johnson, Paice  \&
  Maccarone}{Gandhi et~al.}{2019}]{Ghandi}
Gandhi P.,  Rao A.,  Johnson M. A.~C.,  Paice J.~A.,   Maccarone T.~J.,  2019,
  \mn@doi [Monthly Notices of the Royal Astronomical Society]
  {10.1093/mnras/stz438}, 485, 2642

\bibitem[\protect\citeauthoryear{{Giovannelli, F.}, {Bisnovatyi-Kogan, G. S.}
  \& {Klepnev, A. S.}}{{Giovannelli, F.} et~al.}{2013}]{Giovanelli}
{Giovannelli, F.} {Bisnovatyi-Kogan, G. S.}  {Klepnev, A. S.} 2013, \mn@doi
  [A\&A] {10.1051/0004-6361/201220800}, 560, A1

\bibitem[\protect\citeauthoryear{Henze, Ness, Darnley, Bode, Williams, Shafter,
  Kato  \& Hachisu}{Henze et~al.}{2014}]{Henze_2014}
Henze M.,  Ness J.-U.,  Darnley M.~J.,  Bode M.~F.,  Williams S.~C.,  Shafter
  A.~W.,  Kato M.,   Hachisu I.,  2014, \mn@doi [Astronomy {\&} Astrophysics]
  {10.1051/0004-6361/201423410}, 563, L8

\bibitem[\protect\citeauthoryear{{Howell} \& {Mason}}{{Howell} \&
  {Mason}}{2018}]{Howell}
{Howell} S.~B.,  {Mason} E.,  2018, \mn@doi [\aj] {10.3847/1538-3881/aadd13},
  \href {https://ui.adsabs.harvard.edu/abs/2018AJ....156..198H} {156, 198}

\bibitem[\protect\citeauthoryear{{Ivezi{\'c}} et~al.,}{{Ivezi{\'c}}
  et~al.}{2019}]{Ivezic}
{Ivezi{\'c}} {\v{Z}}.,  et~al., 2019, \mn@doi [\apj]
  {10.3847/1538-4357/ab042c}, \href
  {https://ui.adsabs.harvard.edu/abs/2019ApJ...873..111I} {873, 111}

\bibitem[\protect\citeauthoryear{MacDonald et~al.,}{MacDonald
  et~al.}{2014}]{MacDonald_2014}
MacDonald R. K.~D.,  et~al., 2014, \mn@doi [The Astrophysical Journal]
  {10.1088/0004-637x/784/1/2}, 784, 2

\bibitem[\protect\citeauthoryear{Orosz et~al.,}{Orosz et~al.}{2001}]{Orosz}
Orosz J.~A.,  et~al., 2001, \mn@doi [The Astrophysical Journal]
  {10.1086/321442}, 555, 489

\bibitem[\protect\citeauthoryear{{{\"O}zel}, {Psaltis}, {Narayan}  \&
  {McClintock}}{{{\"O}zel} et~al.}{2010}]{Ozel}
{{\"O}zel} F.,  {Psaltis} D.,  {Narayan} R.,   {McClintock} J.~E.,  2010,
  \mn@doi [\apj] {10.1088/0004-637X/725/2/1918}, \href
  {https://ui.adsabs.harvard.edu/abs/2010ApJ...725.1918O} {725, 1918}

\bibitem[\protect\citeauthoryear{{Patterson}}{{Patterson}}{1984}]{Patterson}
{Patterson} J.,  1984, \mn@doi [\apjs] {10.1086/190940}, \href
  {https://ui.adsabs.harvard.edu/abs/1984ApJS...54..443P} {54, 443}

\bibitem[\protect\citeauthoryear{Raskin, Morren, Pessemier, Bloemen,
  Klein-Wolt, Roelfsema, Groot  \& Aerts}{Raskin et~al.}{2016}]{Raskin_2016}
Raskin G.,  Morren J.,  Pessemier W.,  Bloemen S.,  Klein-Wolt M.,  Roelfsema
  R.,  Groot P.,   Aerts C.,  2016, in Evans C.~J.,  Simard L.,   Takami H.,
  eds, {SPIE} Proceedings. {SPIE}, \mn@doi{10.1117/12.2232485}, \url
  {https://doi.org/10.1117%2F12.2232485}

\bibitem[\protect\citeauthoryear{Rosas-Guevara et~al.,}{Rosas-Guevara
  et~al.}{2015}]{Rosas-Guevara}
Rosas-Guevara Y.~M.,  et~al., 2015, \mn@doi [Monthly Notices of the Royal
  Astronomical Society] {10.1093/mnras/stv2056}, 454, 1038

\bibitem[\protect\citeauthoryear{Russell et~al.,}{Russell
  et~al.}{2019}]{Russell_2019}
Russell D.~M.,  et~al., 2019, \mn@doi [Astronomische Nachrichten]
  {10.1002/asna.201913610}, 340, 278

\bibitem[\protect\citeauthoryear{{Shahbaz} \& {Kuulkers}}{{Shahbaz} \&
  {Kuulkers}}{1998}]{Shahbaz_1998}
{Shahbaz} T.,  {Kuulkers} E.,  1998, \mn@doi [\mnras]
  {10.1046/j.1365-8711.1998.29511221.x}, \href
  {https://ui.adsabs.harvard.edu/abs/1998MNRAS.295L...1S} {295, L1}

\bibitem[\protect\citeauthoryear{Southworth, Copperwheat, Gänsicke  \&
  Pyrzas}{Southworth et~al.}{2010}]{Southworth_2010}
Southworth J.,  Copperwheat C.~M.,  Gänsicke B.~T.,   Pyrzas S.,  2010,
  \mn@doi [Astronomy and Astrophysics] {10.1051/0004-6361/200913576}, 510, A100

\bibitem[\protect\citeauthoryear{Tucker et~al.,}{Tucker et~al.}{2018}]{Tucker}
Tucker M.~A.,  et~al., 2018, \mn@doi [The Astrophysical Journal Letters]
  {10.3847/2041-8213/aae88a}, 867, L9

\bibitem[\protect\citeauthoryear{Warner}{Warner}{1987}]{Warner}
Warner B.,  1987, \mn@doi [Monthly Notices of the Royal Astronomical Society]
  {10.1093/mnras/227.1.23}, 227, 23

\bibitem[\protect\citeauthoryear{Wu, Yu, Li, Maccarone  \& Li}{Wu
  et~al.}{2010}]{Wu_2010}
Wu Y.~X.,  Yu W.,  Li T.~P.,  Maccarone T.~J.,   Li X.~D.,  2010, \mn@doi [The
  Astrophysical Journal] {10.1088/0004-637x/718/2/620}, 718, 620

\bibitem[\protect\citeauthoryear{Yao et~al.,}{Yao et~al.}{2021}]{Yao_2021}
Yao Y.,  et~al., 2021, \mn@doi [The Astrophysical Journal]
  {10.3847/1538-4357/ac15f9}, 920, 120

\bibitem[\protect\citeauthoryear{Zwart, Dewi  \& Maccarone}{Zwart
  et~al.}{2004}]{Portegies_Zwart_2004}
Zwart S. F.~P.,  Dewi J.,   Maccarone T.,  2004, \mn@doi [Monthly Notices of
  the Royal Astronomical Society] {10.1111/j.1365-2966.2004.08327.x}, 355, 413

\bibitem[\protect\citeauthoryear{{van Paradijs} \& {McClintock}}{{van Paradijs}
  \& {McClintock}}{1994}]{vanParadijs}
{van Paradijs} J.,  {McClintock} J.~E.,  1994, \aap, \href
  {https://ui.adsabs.harvard.edu/abs/1994A&A...290..133V} {290, 133}

\makeatother
\end{thebibliography}

\nocite{Giovanelli}
\nocite{Henze_2014}
\nocite{Southworth_2010}
\nocite{Patterson}
\nocite{Raskin_2016}

\begin{table*}

	\centering
	\caption{Apparent \& absolute magnitudes in V-filter unless otherwise noted, color excess factor, distance and period of binary systems}
	
	\label{tab:example_table} 
	\small\addtolength{\tabcolsep}{6pt}
	\begin{tabular}[hbt!]{lrcrcr} 
		\hline
		Object ID & $m$ & $E(B-V)$ $^\mathrm{a}$ & $M$ $^\mathrm{b}$ & $Log(P)$ (hr) $^\mathrm{c}$ & $d$ (kpc)\\
		\hline
4U 1543-475 = IL Lup & 14.9 & 0.50 & -1.03 & 1.43 & 7.5 $\pm$ 0.5 \\
4U 1755-338 = V4134 Sgr & 18.5 & 0.62 & 2.50 & 0.64 & 6.5 $\pm$ 2.5 \\
1H 1659-487 = GX 339-4 = V821 Ara & 14.7 & 1.20 & -2.91 & 1.62 & >6 \\
3A 0620-003 = N Mon 1975 = V616 Mon & 11.1 & 0.35 & -0.09 & 0.89 & 1.1 $\pm$ 0.1 \\
H 1705-250 = N Oph 1977 = V2107 Oph &  $B$\;= 16.3 & 0.50 & 0.08 & 1.09 & 8.6 $\pm$ 2.1 \\
GS 1354-64 = BW Cir &  16.9 & 1.00 & -3.19 & 1.79 & 25 \\
GS 2000+251 = QZ Vul & 16.4 & ~1.40 & -0.09 & 0.92 & 2.7 $\pm$ 0.7 \\
GS 2023+338 = V404 Cyg & 11.7 & 1.30 & -4.22 & 2.19 & 2.4 $\pm$ 0.14 \\
GRS 1124-684 = N Mus 1991 = GU Mus & 13.5 & 0.30 & -1.28 & 1.02 & 5.9 $\pm$ 0.3 \\
GRO J0422+32 = V518 Per &  $R$\;= 12.7 & 0.30 & -0.26 & 0.71 & 2.5 $\pm$ 0.3 \\
GRS 1009-45 = N Vel 1993 = MM Vel & $R$\;= 14.6 & ~0.21 & 1.07 & 0.84 & 3.8 $\pm$ 0.3 \\
GRO J1655-40 = N Sco 1994 = V1033 Sco & $R$\;= 13.6 & 1.30 & -3.01 & 1.80 & 3.2 $\pm$ 0.2 \\
XTE J1550-564  = V381 Nor & 16.6 & 1.33 & -0.79 & 1.57 & 4.5 $\pm$ 0.5 \\
SAX J1819.3-2525 V4641 Sgr & 8.80 & 0.35 & -6.25 & 1.83 & 6.2 $\pm$ 0.7 \\ 
XTE J1859+226= V406 Vul & $R$\;= 15.5 & 0.58 & -2.13 & 0.82 & 12.5 $\pm$ 1.5 \\ 
XTE J1118+480 = KV UMa & 13.0 & 0.02 & 1.75 & 0.61 & 1.7 $\pm$ 0.1 \\
XTE J1650-500 & $B$\;= 16.8 & 1.50 & 0.08 & 0.89 & 2.6 $\pm$ 0.7 \\
SWIFT J1753.5-0127 & $R$\;= 15.9 & 0.45 & 3.89 & 0.51 & 6 $\pm$ 2 \\
XTE J1752-223 V5678 Sgr & $J$\;= 15.8  & 1.40 & -2.49 & 0.84 & 6 $\pm$ 2 \\ 
MAXI J1659-152 V2862 Oph &  16.78 & 0.34 & 1.03 & 0.38 & 8.6 $\pm$ 3.7 \\
MAXI J1836-194 & 16.33 & 0.60 & 0.24 & 0.68 & 7 $\pm$ 3 \\
MAXI J1820+070 Asassn-18ey = V3721 Oph & 14.0 & 0.18 & 1.09 & 1.22 & 2.9 $\pm$ 0.3 \\

\hline
\multicolumn{5}{l}{$^a$ Correction factors from Corral-Santana et al. (2016).}\\
\multicolumn{6}{l}{\footnotesize$^b$ Derived using $E(B-V)$ value and eq. (1).}\\
\multicolumn{6}{l}{\footnotesize$^c$ Logarithms of $P$ from Corral-Santana (2016).}\\

\end{tabular}
\end{table*}

\newpage
\begin{figure*}
\centering
	\includegraphics[width=2.2
	\columnwidth]{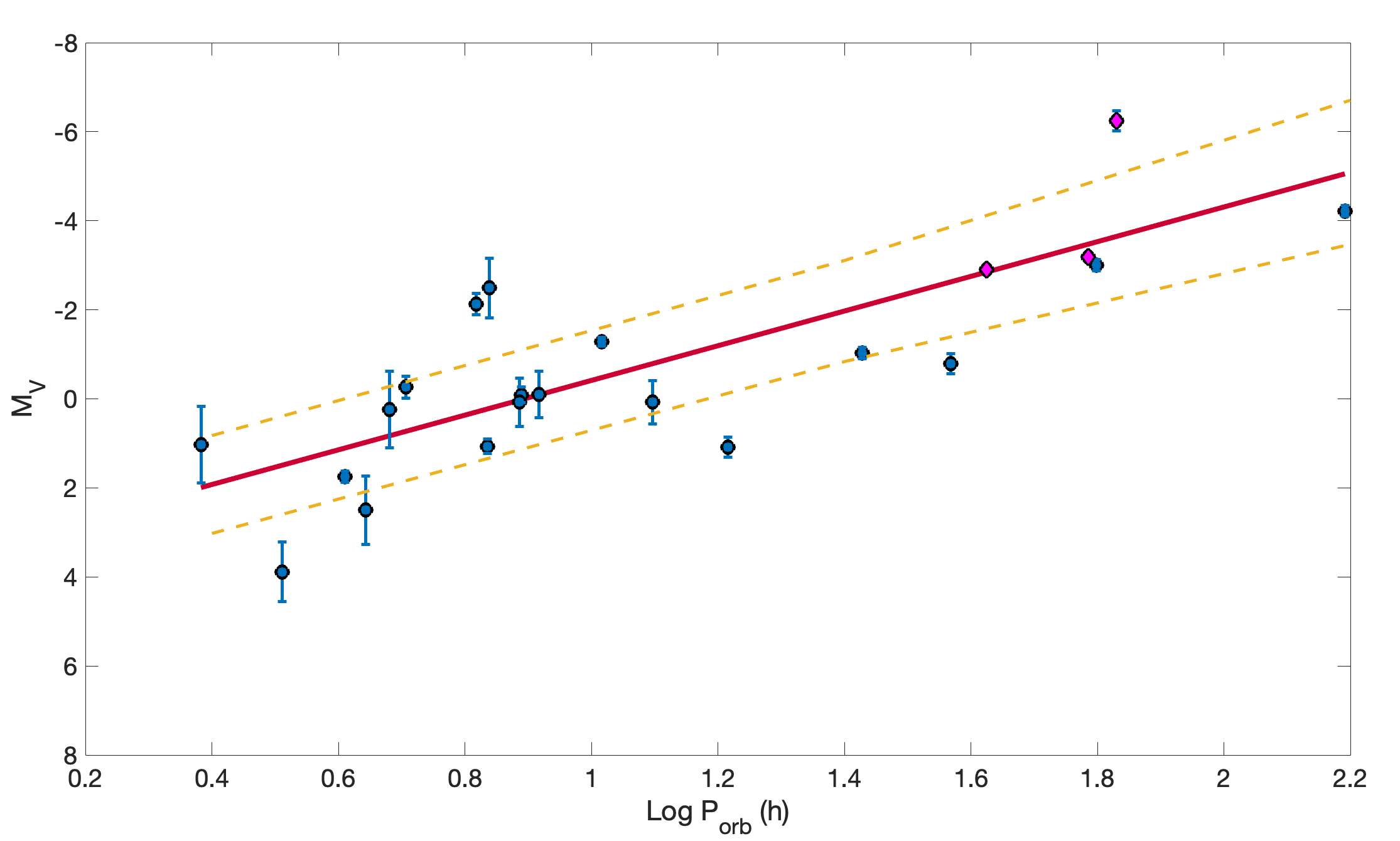}
    
	\includegraphics[width=2.2
	\columnwidth]{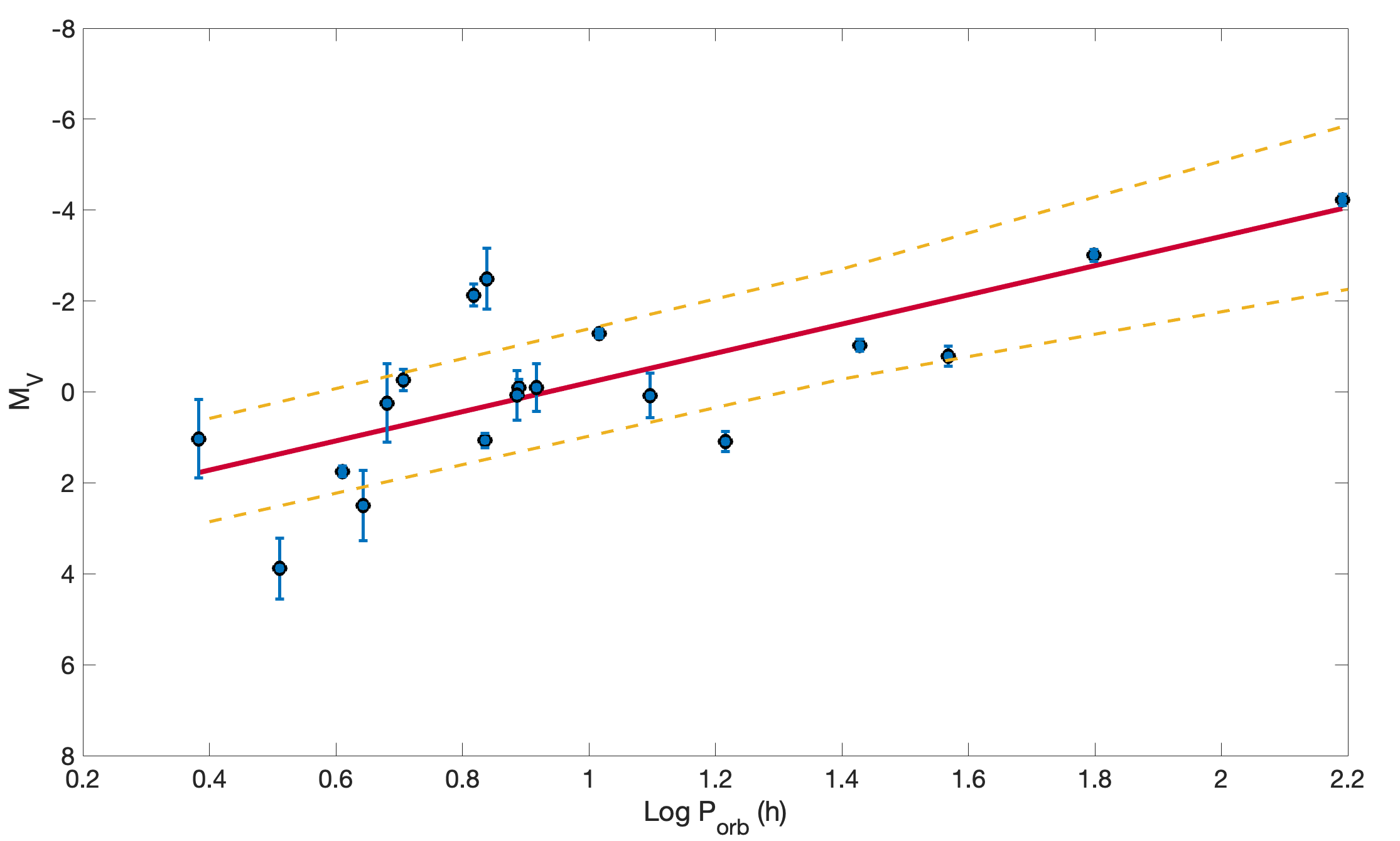}
    \caption{Top: The relation between $M_V$ and $LogP_{orb}$ for the sample with errorbars corresponding to individual distance error. Data includes BW~Cir, V821~Ara, and outlier V4641~Sgr indicated by magenta markers. Dashed gold lines indicate the 68\% confidence interval. Bottom: The relation between $M_V$ and $LogP_{orb}$ excluding BW~Cir, V821~Ara and V4641~Sgr.}
    \label{fig: example_figure}
\end{figure*}


\bsp	

\label{lastpage}
\end{document}